\documentclass[pre,twocolumn,groupedaddress,showpacs,pdfoutput=1]{revtex4-1}
\usepackage{graphicx}
\usepackage{epsfig}
\usepackage{amssymb,amsmath,amsfonts,hyperref}
\usepackage{wasysym}
\usepackage{bm}
\usepackage{latexsym}
\usepackage{eucal}
\usepackage{verbatim}
\usepackage[normalem]{ulem}
\usepackage{color}

\newcommand{\be}{\begin{equation}}
\newcommand{\ee}{\end{equation}}

\newcommand{\bea}{\begin{eqnarray}}
\newcommand{\eea}{\end{eqnarray}}

\begin{document}

\title{Indicators of Conformal Field Theory: \\
entanglement entropy and multiple-point correlators}
\author{Pranay Patil}
\author{Ying Tang}
\author{Emanuel Katz}
\author{Anders W. Sandvik}

\affiliation{Department of Physics, Boston University, 590 Commonwealth Avenue, Boston, Massachusetts 02215, USA}

\begin{abstract}
The entanglement entropy (EE) of quantum systems is often used as a test of low-energy descriptions by conformal 
field theory (CFT). Here we point out that this is not a reliable indicator, as the EE often shows the same behavior even 
when a CFT description is not correct (as long as the system is asymptotically scale-invariant). We use constraints on the 
scaling dimension given by the CFT with SU(2) symmetry to provide alternative tests with two- and four-point correlation
functions, showing examples for quantum spin models in 1+1 dimensions. In the case of a critical amplitude-product state 
expressed in the valence-bond basis (where the amplitudes decay as a power law of the bond length and the wave function
is the product of all bond amplitudes), we show that even 
though the EE exhibits the expected CFT behavior, there is no CFT description of this state. We provide numerical tests of 
the  behavior predicted by CFT for the correlation functions in the critical transverse-field Ising chain and the $J$-$Q$ spin 
chain, where the conformal structure is well understood. That behavior is not reproduced in the amplitude-product state.
\end{abstract}

\maketitle

\section{Introduction}

The emergence of conformal invariance at a quantum phase transition is helpful for understanding many aspects of the 
critical behavior, including determination of the critical exponents. Thus, knowledge that a quantum phase transition 
has conformal invariance is quite useful and it is important that we have reliable methods to determine the same. The 
entanglement entropy (EE) was predicted to have distinct behaviors which would reflect the central charge of the corresponding
conformal field theory (CFT) \cite{Cardy}, and as such it has often been taken to be an indicator of conformal invariance. 
Later it was shown numerically \cite{Pochung} and analytically \cite{RM,PR2,PencPollmann} that in some cases the behavior seen for the 
EE was not exclusive to conformal invariance and could be seen in some systems without conformal invariance as well. In fact,
the derivation of the EE scaling form does not rely on conformal invariance, but only on the less constraining scale invariance \cite{Cardy}, 
though the number associated with the conformal charge of course has that meaning only for a CFT. In contrast, the two-point correlation 
functions have a scaling behavior and functional form which are both completely controlled by the CFT. These can serve as litmus tests for 
an underlying  CFT.

Here we will present a simple example of a quantum spin chain whose entanglement entropy shows the dependence on the size of 
the entangled region as predicted by a CFT and yet is not described by a CFT; we prove the latter using two- and four-point spin correlation functions. 
A previous work \cite{RM} by Rafael and Moore shows the possibility of a universal scaling of the entanglement entropy in the case 
of random spin chains which would not be expected to be conformally invariant. Here we demonstrate numercially the same in a system 
without randomness, namely a critical amplitude-product state (APS) on a spin-$1/2$ chain. The APS is an SU(2) symmetric superposition 
state made out of all pair-singlet covers of the 1D bipartite lattice. This state can be tuned to criticality for particular weightings 
of the singlet covers which contribute to it, and one may then ask whether a state similar to the ground state of the Heisenberg 
Hamiltonian could be constructed this way. We here compute the EE and correlation functions at such critical points and conduct
CFT tests. While the EE exhibits the expected scaling behavior, the correlation functions do not. It should be noted that here we do not 
know the underlying Hamiltonians for which the APS states are ground states, though such parent Hamiltonians should exist. To contrast 
the negative CFT tests for the APS with results for actual conformally-invariant systems, we will also study critical points of two Hamiltonians; 
the transverse-field Ising model (TFIM) on a chain and the $J$-$Q$ chain (the Heisenberg chain supplemented by a four-spin
interaction). The latter hosts a phase transition between the quasi-ordered (critical) state similar to that of the Heisenberg
chain and a dimerized phase similar to that in the Heisenberg chain with frustrated interactions.

In Sec.~\ref{sec2} we will briefly review the known analytical results for two-dimensional CFTs restricted by SU(2) invariance. 
In Sec.~\ref{sec3a} we develop numerical tests of CFT in the simpler setting of the Ising model, using Monte Carlo and quantum Monte Carlo
(QMC) results for correlation functions of the classical 2D model as well as the TFIM chain. In Sec.~\ref{sec3b} we move on to QMC results 
for the SU($2$) symmetric $J$-$Q$ chain. In Sec.~\ref{sec4}, we will introduce the APS in detail and show its cosistency with CFT as 
far as the EE is concerned, and inconsistencies with other CFT predictions. We give a brief summary and conclusions in Sec.~\ref{sec:concl}.

\section{\label{sec2}Results from CFT}

A conformal field theory contains primary operators whose correlation functions are constrained due to the conformal 
invariance condition \cite{241}. To establish the setting of our study and introduce definitions, we here give a brief
summary of correlation functions within CFT.

It has been shown \cite{A,HA} that quantum spin chains with a certain set of interactions 
allowed by spin and translation symmetries have conserved currents which can be recreated by a scalar field theory; in this 
case the $k=1$ Wess-Zumino-Witten (WZW) model. The WZW model has a conformal symmetry as well as a SU(2) symmetry. Using the
conformal structure, one can predict that the correlation functions have functional forms controlled by the operator product
expansion in the CFT. The two symmetries lead to two currents that follow a combined Virasoro Kac-Moody algebra \cite{247} 
that controls the properties of the primary fields making up the CFT.

The $k$=1 WZW Lagrangian is written in terms of a spinless primary field $g$, which has scaling dimension $\Delta=1/2$, as can be seen by 
using the algebra and the operator product expansion of $g$ with the SU(2) current \cite{Fran}. These operator product expansions 
lead to Ward identities \cite{247}, which are differential equations obeyed by the correlation functions of primary operators. 
The equations force correlation functions to follow power-law behavior where the exponents are controlled by the scaling dimensions
of the primary operators it is made up of. Recalling that $g$ itself is a $2\times 2$ matrix, we expect the correlation functions to be 
matrices themselves. In our numerical investigations we will make use of the two-point correlation function, which
has the expected asymptotic form given by
\begin{equation}
\label{eq4}
\langle g(z,\bar{z})g(0,0)\rangle \sim z^{\Delta}\bar{z}^{\Delta} M_{4},
\end{equation} 
and the four-point function
\begin{eqnarray}
&&\langle g(z_1,\bar{z_1})g(z_2,\bar{z_2})g(z_3,\bar{z_3})g(z_4,\bar{z_4})\rangle \label{eq5} \\
&&\sim [(z_1-z_4)(z_2-z_3)(\bar{z_1}-\bar{z_4})(\bar{z_2}-\bar{z_3})]^{-\Delta} F(p,\bar{p}) M_{16}, \nonumber
\end{eqnarray}
where $z$ lives on the infinite complex plane, $M_{n}$ is a short-hand for an $n\times n$ matrix and
we define the set of anharmonic coefficients $(p,\bar{p})$ as \cite{247}
\begin{equation}
p=\frac{(z_1-z_2)(z_3-z_4)}{(z_1-z_4)(z_3-z_2)},\ \  \bar{p}=\frac{(\bar{z_1}-\bar{z_2})(\bar{z_3}-\bar{z_4})}{(\bar{z_1}-\bar{z_4})(\bar{z_3}-\bar{z_2})}.
\end{equation}
Through the numerics, we will observe the behavior of the four-point correlation function for fixed anharmonic ratios.
The structure of the matrix $M$ is controlled by selection rules given by the SU(2)$\times$SU(2) decomposition that it follows \cite{247}.
This will not be important in our case, as we will be dealing with correlators of ${\rm tr}(g)$, and the matrix will be traced over 
with some constraints, but this will not affect the scaling behavior.

The $k$=1 WZW model can be connected to spin chains by looking at the Heisenberg model at criticality and the additional interactions
allowed \cite{A}. From a numerical standpoint, we study the $J$-$Q$ chain, which has the same critical behavior as the Heisenberg 
chain \cite{TanSan} and can be tuned through a quantum phase transition into a dimerized state. Similar to the case of the frustrated
Heisenberg chain \cite{eggert96}, at the dimerization point the logarithmic corrections existing in the system up until this point vanish, 
thus removing the small but visible violations of conformal invariance. For simplicity we here discuss the theoretical results within 
the Heisenberg Hamiltonian, given by
\begin{equation}\label{eq6}
H=\sum_{i=1}^N {\bf S}_i \cdot {\bf S}_{i+1},
\end{equation}
where ${\bf S}_i$ are the standard spin-$1/2$ operators, and later carry them
over into the $J$-$Q$ chain. The spin operators
can be written in terms of standard fermion creation and annihilation
operators \cite{A}, which can then be expanded in the Fourier modes
close to the Fermi surface. Following the work by Affleck \cite{A},
the Fourier-mode fermionic operators can be written in terms of
currents and $g$ field operators of the WZW theory. The identification
of the fermionic operators with $g$ is done by noting that the currents
in both models have the same commutation relations \cite{247},
thus implying that they can be used interchangeably. Coarse
graining the spin chain so that the current $J$ and field $g$ can
be treated as smooth objects, we can write the spin operator as
\begin{equation}\label{eq7}
S^b_n=a[J^b+\bar{J}^b+(-1)^n{\rm tr}[(g+g^{\dagger})\tau^b)],
\end{equation}
where $b$ is one of $\{x,y,z\}$, $n$ is the lattice site number
and $\tau^b$ is the correspoding Pauli matrix here.
The scaling dimension of $J$ is unity as it is a conserved current
and that of the $g$ field is $1/2$ as mentioned earlier. Using these observations one can predict 
that the two-point spin correlation should scale with exponent $1$ to leading order in $1/L$,
the three-point one should scale with the exponent of $3/2$, etc. 

In our case we will only use equal-time correlation functions, motivated primarily
by the fact that the APS is only given as a wave-function, with no Hamiltonian to generate time correlations. This is not an
uncommon situation when exploring various quantum states. Nevertheless, as mentioned above, in principle there still exists
a parent Hamiltonian for a given wave function, and the constraints of the CFT allow for tests for a CFT by considering 
solely equal-time correlations. 

Possible equal-time correlation functions for two and three spins at different 
sites on a chain are given by
\begin{equation}\label{eq8}
O^{ab}_{ij}=\langle S_i^aS_j^b\rangle,~~~~O^{abc}_{ijk}=\langle S_i^aS_j^bS_k^c\rangle,
\end{equation}
where $a,b,c$ each are one of $\{x,y,z\}$. These correlation functions are similar to the ones mentioned above
and can be written in terms of the Wess-Zumino $g$ field. Note, however, that unless $a=b$ the two-point correlator vanishes
in the ground states of the SU($2$) symmetric Hamiltonians we will consider, while the three-point correlator vanishes for any
choise of the components $a,b,c$. For models and combinations of the components where the correlators do not vanish,
the general conformal contraints discussed above forces them to be of the form
\begin{equation}\label{c3g}
\langle O_iO_j\rangle \sim \frac{1}{z_{ij}^{2\Lambda_i}},~~~~ 
\langle O_iO_jO_k\rangle \sim \frac{1}{z_{ij}^{\Delta_{ij}}z_{jk}^{\Delta_{jk}}z_{ik}^{\Delta_{ik}}},
\end{equation}
where $z_{ij}=|z_i-z_j|$ is the spacetime separation (in our case purely spatial separation) between the
operators on site $i$ and $j$, and $\Delta_{12}=\Lambda_1+\Lambda_2-\Lambda_3$ and so on. The lattice spin 
correlation functions approximate to the continuum functions constructed out of the fields of the CFT; in the case 
of the spin correlations we will denote the corresponding field by $\bm{\sigma}$. 

Averaging over the spin components, we will study the two-point correlator 
\begin{equation}\label{c2s}
C_2(x_1)=\langle {\bf S}_0\cdot {\bf S}_{x_1}\rangle,
\end{equation}
where $0$ is a reference site and $x_1$ the distance to it. Since the three-point correlator vanishes for the SU($2$) 
Hamiltonians, when considering those we will instead study the rotationally-averaged four-point correlator
\begin{equation}\label{c4}
C_4(x_1,x_2,x_3)=\langle ({\bf S}_0\cdot {\bf S}_{x_1})({\bf S}_{x_2}\cdot {\bf S}_{x_3})\rangle,
\end{equation}
for which scaling forms generalizing Eqs.~(\ref{c3g}) will apply. For convenience in 
finite-size scaling, we will often choose all distances to be fractions of the system size $L$. 

The global group of a CFT, which consists of translations, rotations, scaling and conformal
transformations, restricts the scaling forms of correlation functions to Eq.~\eqref{c3g}
and the full Virasoro algebra forces these functions to depend only on the conformal distance \cite{Fran}. 
For all the 1D systems we study in this work,
we will use periodic boundary conditions, leading to a cylindrical spacetime geometry that can be mapped to the complex plane 
through a conformal transformation ($x\to\theta,t\to r$) \cite{Cardy}. The conformal equal-time distances on the complex plane 
correspond to  spatial distances on the lattice through \cite{Fran} 
\begin{equation}\label{confdist}
\Delta z=L\sin\left (\pi\frac{\Delta x}{L}\right ). 
\end{equation}
Thus, the two-point function on periodic chains, which scales with twice the spin dimension $\Lambda_{\sigma}$ 
[which is $1/2$ for the SU(2) case], should take the form
\begin{equation}\label{c2}
C_2(x_1) \propto \bigg(\frac{1}{L\sin(\pi x_1/L)}\bigg)^{2\Lambda_{\sigma}},
\end{equation}
for sufficiently large distance $x_1$. The four-point function should also
take a similar form where all the $z_{ij}$'s are replaced by the conformal
distance.

The global group of a CFT is invariant in all dimensions and the
scaling constraints it sets on correlation functions can be easily seen
in finite-size scaling. For the SU(2) symmetry, we were not able to construct
a three-point correlator as the only field operator we could test
was $\bm{\sigma}$.
We would expect any quantum spin chain with a Hamiltonian which permits the 
full SU(2)$\times$SU(2) symmetry 
(coming from decoupling of left and right moving currents)
in its interaction terms
to be described by the $k=1$ WZW theory. This would imply
that the correlation functions in the APS, whose unknown parent Hamiltonian
must have SU(2) symmetry due to the states being made out of singlets,
would have the same scaling dimensions as predicted for the $J$-$Q$ model
if it had the correct interaction terms.
It is known \cite{HA} that given a quantum spin chain which
is translationally invariant against a shift of 1 site,
the only local interaction which can push the system away from
a SU(2)$\times$SU(2) symmetry is a negative coupling in the term
coupling left and right currents (${\bf J}_L\cdot {\bf J}_R$).
This does not happen in the Heisenberg chain \cite{HA} as the coupling
starts off positive and is thus irrelevant, and we also would not expect 
it in the $J$-$Q$ chain before its dimerization transition. 
The situation is less clear for the APS, where we do not
know the parent Hamiltonian, which, apart from the potential attractive local interactions
mentioned above, may contain long-range interactions that ruin the CFT description.

\section{\label{sec3a}Classical and quantum Ising models}

Though the main focus of this paper is on systems with SU($2$) symmetry, we make a brief detour to systems without this symmetry,
namely the classical 2D Ising model and the transverse-field Ising chain. We develop and test practical procedures to detect the signatures of 
the two- and three-point correlators predicted by a CFT.

The critical-point TFIM Hamiltonian is
\begin{equation}\label{eqHI}
H=-\sum_{i=1}^N \sigma^z_i\sigma^z_{i+1} - \sum_{i=1}^N \sigma^x_i.
\end{equation}
This Hamiltonian can be written in terms of Majorana fermions in the continuum as follows \cite{I2DCFT}:
\begin{equation} 
H=\frac{1}{2}\Big(-i\frac{\partial \psi_R(x)}{\partial x}\psi_R(x)\Big)-
\frac{1}{2}\Big(-i\frac{\partial \psi_L(x)}{\partial x}\psi_L(x)\Big).
\end{equation}
In general this Hamiltonian should also have a mass term, but it  disappears at criticality,
leaving $H$ invariant under scaling. From the Hamiltonian we see that 
the equations of motion imply that the two fermions are completely disconnected (due to the absence of a mass term) and thus it is
sufficient to study just one of them.

The spin operators in which the Hamiltonian of Eq.~\eqref{eqHI} is expressed are found to be nonlocal in the majorana fermions \cite{I2DCFT}. 
With periodic boundary conditions on the majorana fermions, it was found that a possible set of non-trivial primary operators of the CFT
could be $\sigma$,$\mu$,$\epsilon$ which are the order, disorder, and energy density operators, respectively \cite{I2DCFT}.
As discussed in the previous section, the correlation functions of these primary operators are constrained by the operator product expansion in a CFT \cite{241} to power laws 
which are controlled by the scaling dimensions of the operators as in Eq.~\eqref{c3g}. The scaling dimensions for the primary operators are \cite{241}
\begin{eqnarray} 
\Delta_{\sigma}=\frac{1}{8},~~
\Delta_{\mu}=\frac{1}{8},~~
\Delta_{\epsilon}=1.
\end{eqnarray}
Using these and the forms predicted by the CFT, Eq.~\eqref{c3g}, we expect the following connected two-point and three-point 
functions to scale as
\begin{equation}\label{corrsc}
C_s(aL)=\langle\sigma(0)\sigma(aL)\rangle_c\sim L^{-\frac{1}{4}},
\end{equation}
\begin{equation}
\label{corrsc2}
C_e(aL)=\langle\epsilon(0)\epsilon(aL)\rangle_c\sim L^{-2},
\end{equation}
\begin{equation}
\label{corrsc3}
C_3(aL,bL)=\langle\sigma(0)\epsilon(aL)\sigma(bL)\rangle_c\sim L^{\frac{3}{4}}L^{-1}L^{-1},
\end{equation}
where $a,b\in(0,1)$ and the functional dependence on $a,b$ depends on the boundary conditions being used; an example of which is 
provided in Eq.~\eqref{c2}. It is worth noting that observation of scaling forms such as (\ref{corrsc})-(\ref{corrsc3})
do not in themselves guarantee a CFT as fields outside of a CFT can also have well defined scaling dimensions, while
a detailed form of the dependence on the conformal distance similar to Eq.~\eqref{c2} is more specific to a CFT.

\begin{figure}[t]
\includegraphics[width=\hsize]{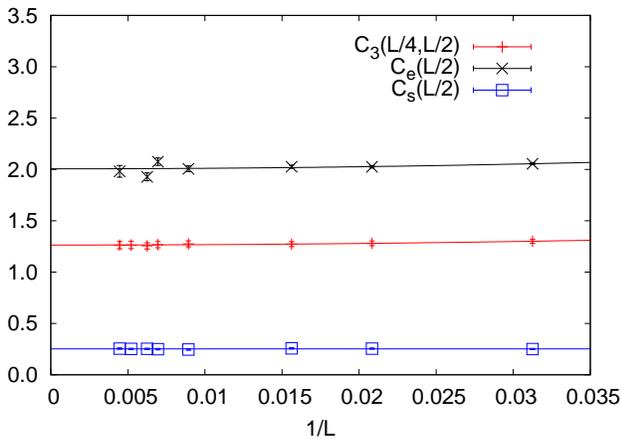}
\caption{Two- and three-point functions for the classical critical 2D Ising model.
The exponents for
$\langle\sigma\sigma\rangle$, $\langle\epsilon\epsilon\rangle$,
and $\langle\sigma\sigma\epsilon\rangle$ are found to be $0.252(4)$, $2.01(3)$, and $1.263(5)$
respectively, where the numbers within parenthesis indicate the statistical error
(one standard deviation) of the preceding digit. The curves show fits to Eq.~(\ref{gammalform}) using $b=2$.}
\label{Fig3}
\end{figure}

We would expect to see these CFT behaviors in both the 2D classical Ising model and the transverse-field quantum Ising chain, as 
they are connected through the transfer-matrix approach. On the lattice, $\sigma(i)$ corresponds to the spin at location $i$ and
$\epsilon(i)$ corresponds to the Ising part of the energy density, $\sigma(i)\sigma(i+1)$.  For the TFIM, the points which 
are used to observe the correlation functions are picked along the spin chain at equal time, whereas in the 2D classical 
case we expect emergent rotational invariance for large separation and see the same correlation functions should be observed
provided the distances are the same. 

Note that the conformal-distance form (\ref{c2}) of the two-point function is not applicable to the 2D classical 
Ising model, as our simulations use $L\times L$ lattices with periodic boundary conditions in both dimensions, 
in which case the conformal distance assumes a more complicated form \cite{Cardy}. For the 1D TFIM this is not an issue 
as we use projector QMC \cite{AWTFIM,CWLiu} to obtain results at temperature $T \to 0$, which corresponds to the time 
dimension being much larger (effectively infinite) than the space dimension and the mapping leading to the conformal 
distance of Eq.~\eqref{confdist} is applicable. In principle this form could also be used for the classical model
if $L_x\times L_y$ lattices are used with a very large aspect ratio $L_y/L_x$.

We begin with results for the classical 2D model, for which we have used the Swendsen-Wang cluster Monte Carlo algorithm 
\cite{SW}. Though the critical temperature $T_c$ is known exactly in this case, in order to test procedures where this 
is not the case we here test a method that works also if $T_c$ is not known. We employ the method of flowing 
Binder cumulant crossing points \cite{Luck} to extract exponents for the critical correlation functions, collecting data 
for the Binder cumulant and all the correlations we are interested in over a range of system sizes and temperatures close to
the known critical temperature. For a pair of system sizes $(L,2L)$, we find the crossing point of the Binder cumulants, 
and we can regard this as a definition of the critical temperature for size $L$. (This procedure is illustrated in the 
case of the APS further below in Sec.~\ref{sec4}.) 

\begin{figure}[t]
\includegraphics[width=\hsize]{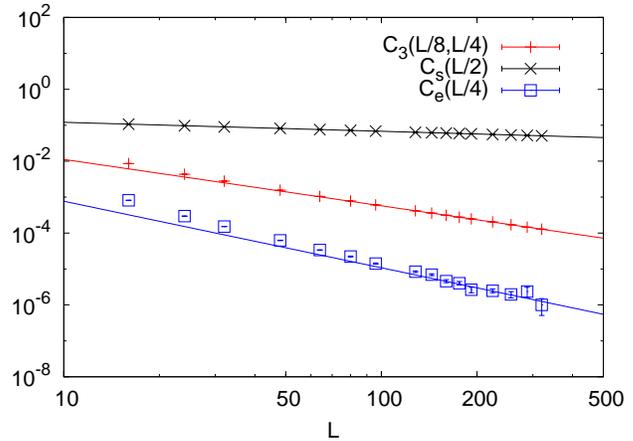}
\caption{Power-law behavior of correlation functions in the 1D TFIM chain at
criticality. The exponents for the spin correlation, energy density
correlation and three-point correlation are $0.2498(2)$,
$1.9(4)$ and $1.29(1)$ respectively.}
\label{Fig4}
\end{figure}

We extract the correlators at the crossing temperature by interpolating within the data set, and extract the floating exponents 
from the two so obtained values by  assuming a power-law dependence in $L$. This process gives us an estimate of the exponents 
at size $L$, and we repeat the same for different size pairs to obtain a series of exponent values that should converge to the 
correct exponents in the thermodynamic limit. The finite-size behavior for an exponent $\gamma$ is expected to be of the form
\begin{equation}
\gamma(L) = \gamma_0 + a\Big(\frac{1}{L}\Big)^b,
\label{gammalform}
\end{equation}
where $\gamma_0$ is the value that we conclude to be the final exponent value and $b$ is an exponent originating from an 
irrelevant field (or, in some cases, from the size corrections to the non-singular part of the free energy).
Our results are presented in Fig.~\ref{Fig3}, and the extracted exponents (listed in the figure caption) are seen to agree 
very well with the expected values. We have used distances proportional to the system size $L$ and fit the flow of the exponents 
to Eq.~\eqref{gammalform} using the leading correction exponent $b=2$ for the Ising universality class.

\begin{figure}[t]
\includegraphics[width=\hsize]{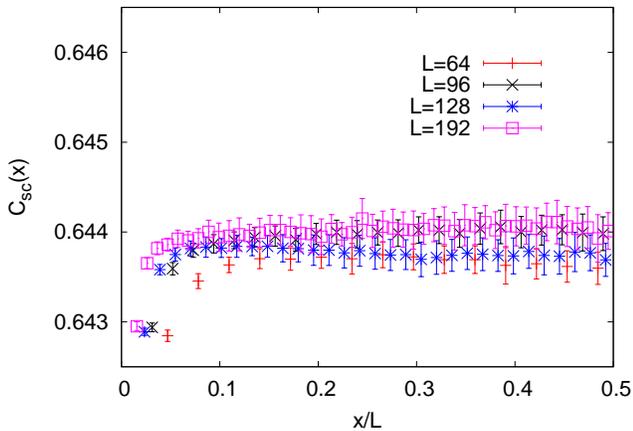}
\caption{Scaled spin correlation function, Eq.~(\ref{sc}), for the critical TFIM on chains of different lengths 
($x/L=0.5$ is the largest separation on a periodic chain). Here we have used $\Delta_{\sigma}=1/2$ as that is the 
scaling dimension of the Ising spin operator \cite{247}.}
\label{Fig5}
\end{figure}

Turning now to the quantum TFIM, we run projector QMC simulations \cite{AWTFIM,CWLiu} to study the ground state 
of the critical Hamiltonian from Eq.~\eqref{eqHI} and extract exponents using different sizes, as illustrated in 
Fig.~\ref{Fig4}. Here we have only carried out simulations exactly at the critical point and we test fitting without
corrections in this case. Since we expect the larger sizes to show better agreement to the asymptotic power-law forms 
than smaller sizes, due to corrections, we fit to a power-law starting with the largest available sizes and add smaller 
sizes as long as the $\chi^2$ value of the fit is acceptable. Once we find the largest set of points with a statistically
acceptable $\chi^2$ value, we drop the smallest size from this set and use the remaining set for our fits. These fits
are presented in Fig.~\ref{Fig4}, and the so obtained exponent values (listed in the figure caption) 
match well the predictions from Eq.~\eqref{corrsc}.

It is important to note here again that through Fig.~\ref{Fig3} and Fig.~\ref{Fig4} we have only proved the scaling behavior 
of the two- and three-point correlator, not the conformal behavior. To study the exact functional form and establish the CFT 
form Eq.~\eqref{c3g}, we need to investigate their dependence on both the point separation and system size and analyze
according to the form \eqref{c2}. To make the functional dependence expressed there more explicit in graphs, we define 
the correlator scaled by the conformal distance as
\begin{equation}\label{sc}
C_{sc}(x)=\left(L\sin\Big(\pi\frac{x}{L}\Big)\right)^{2\Lambda_{\sigma}}C_2(x),
\end{equation}
where $C_2(x)$ is the original correlation function measured at lattice distance $x$. If $C_2(x)$ follows Eq.~\eqref{c2}, the 
scaled correlation should be a constant with respect to the distance $x$. As shown in Fig.~\ref{Fig5}, we find that the scaled 
spin correlation function indeed converges to a constant, even for relatively small system sizes (indicating small scaling 
corrections in this case), and to good precision except for $x/L\approx 0$ (and signs of convergence even when $x \to 0$
are seen when the system size grows). Thus, as expected the TFIM passes the CFT test.

\section{\label{sec3b}J-Q chain}

\begin{figure}[t]
\includegraphics[width=\hsize]{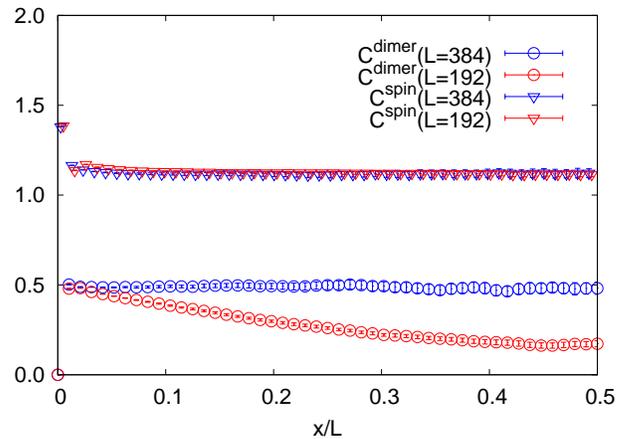}
\caption{The spin and dimer two-point correlation functions of the critical $J$-$Q$ chain scaled with the conformal distance according to Eq.~\eqref{sc}.}
\label{Fig1}
\end{figure}

To test numerical signatures of the $k=1$ WZW model, we study the $J$-$Q$ chain \cite{TanSan} which, in the simplest incarnation, 
is the Heisenberg chain with a particular four-spin interaction added. The Hamiltonian is
\begin{equation}\label{jq2}
H=-J\sum_{i}P_{i,i+1}-Q\sum_{i}P_{i,i+1}P_{i+2,i+3},
\end{equation}
where $P_{i,j}=1/4-{\bf S}_i \cdot {\bf S}_j$ is the singlet projection operator between the sites $i$ and $j$. For $Q/J \to 0$ 
we recover the Heisenberg chain, whereas for $Q/J\gg1$ the four-spin interaction term forces the chain into a valence-bond 
solid state with spontaneously broken translational symmetry. The transition point $Q/J\approx0.8483$ is known from previous 
studies \cite{TanSan,sanyal2011}, and at this point the logarithmic corrections present for smaller coupling ratios vanish. 
All our simulations are done at this critical point, using a projector QMC method formulated in the valence-bond basis to 
extract the properties in the ground state \cite{AWQMC}. The two-point function, $C_2(x)$, has already been found to have 
a scaling exponent of unity \cite{TanSan} as expected of the underlying CFT. 

\begin{figure}[t]
\includegraphics[width=\hsize]{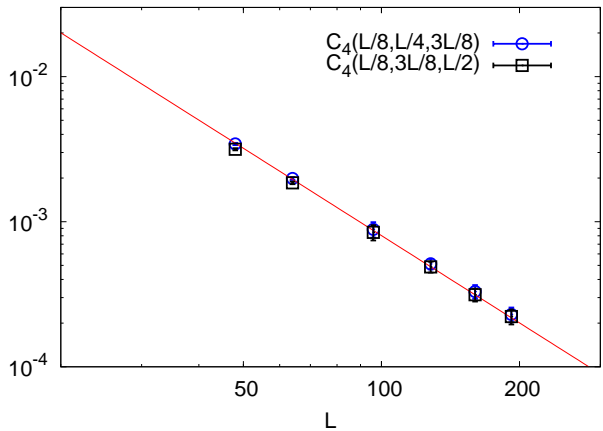}
\caption{Four-point correlators of the critical $J$-$Q$ chain along with a fit 
to the form expected for scaling dimension $2$, shown here for two different 
combinations of cross-ratios and 
system sizes. The prefactor for the two fits is almost the same.}
\label{Fig2}
\end{figure}

In the $J$-$Q$ chain, we have used the two-point correlators for the spin and dimer operators,
which act as order parameters for the critical and dimerized phases, respectively. On the lattice, the dimer operator is 
defined as $D_i={\bf S}_i\cdot {\bf S}_{i+1}$ and its two-point scaled connected correlator approaches a constant when 
the system size is sufficiently large, as does the spin correlator; see Fig.~\ref{Fig1}.

To check the scaling behavior of the four-point function, as defined in Eq.~\eqref{c4}, we plot its finite-size scaled form 
and fit it assuming the scaling dimension $2$, as discussed in Sec.~\ref{sec2}.  The prediction for the four-point function 
also tells us that using different cross ratios should give us the same scaling. Examples of this are shown in Fig.~\ref{Fig2}. 
In cases where we do not know the scaling dimension of the operator making up the correlator, we can use the four-point 
function to determine it, as we  will do for the APS in Sec.~\ref{sec4}.

\section{\label{sec4}Amplitude-Product State}

\begin{figure}[t]
\includegraphics[width=\hsize]{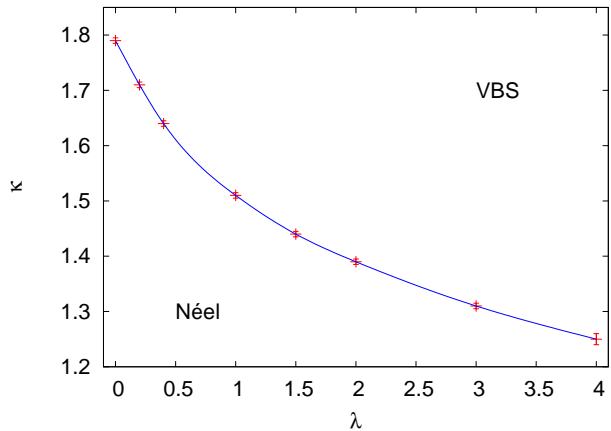}
\caption{Phase diagram of the APS with a critical curve separating the
N\'eel phase and the VBS phase; this matches well the phase diagram previously reported 
in Ref.~\cite{Lin2012}.}
\label{Fig6a}
\end{figure}

The APS is a wavefunction of the resonating valence bond type on a bipartite lattice \cite{Lin2012,LDA}. 
Here we define it in 1D, using parameters $\lambda$ and $\kappa$, as a 
superposition of valence-bond states as,
\begin{equation}\label{eq1}
|\Psi(\lambda,\kappa)\rangle = \sum_{i} A_i(\lambda,\kappa)|V_i\rangle,
\end{equation}
where $|V_i\rangle$ is a tiling of $N_b=L/2$ two-spin singlets between the A-sites and the B-sites on the 1D periodic chain,
and it contributes with a weight $A_i$ to the APS. The weight is given by an amplitude-product depending on the bond-lengths 
present in $|V_i\rangle$;
\begin{equation}\label{eq2}
A_i=\prod_{j=1}^{N_{b}} h(d_j),~~~~
h(d)=\begin{cases} \lambda,\ d=1, \\
d^{-\kappa},\ d>1. \end{cases}
\end{equation}
It should be noted that the asymptotic properties of the state are not
controlled only by the asymptotic power law, as might be naively expected, but also by the short-length weights which are important in this
regard \cite{Beach2009}. 

\begin{figure}[t]
\includegraphics[width=\hsize]{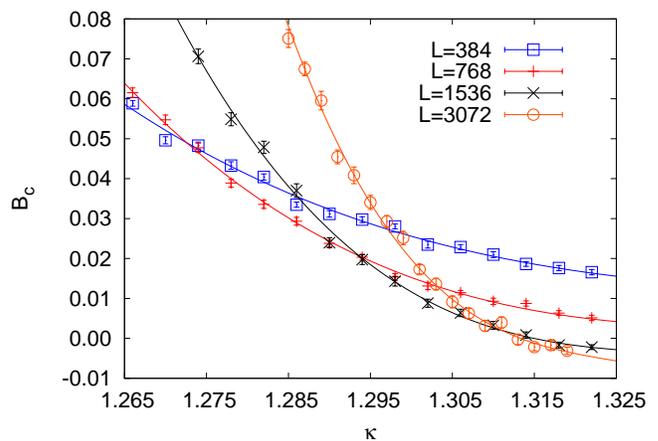}
\caption{Flow of the Binder cumulant of the APS with $\kappa$ for different system sizes at $\lambda=3$. 
We see here the crossing points for three sets of $(L,2L)$ 
(example of extrapolation to critical point shown in inset of Fig.~\ref{Fig7})
and also that $B_c$ 
becomes negative for large sizes 
after the crossing point.}
\label{Fig6}
\end{figure}

We have here chosen a very simple parametrization that allows us to study the effects of both asymptotic and
short-length amplitudes, as in Refs.~\cite{Lin2012,Beach2009}. 
For $\kappa\to 0$ and $\lambda\approx 1$ all valence-bond coverings 
contribute equally and we obtain a long-range ordered N\'eel state,  
whereas for $\kappa\to\infty$ we have only two contributing components, 
each with short bonds on alternating spin pairs, i.e., two degenerate dimerized states. 
In $(\lambda,\kappa)$ space we have a critical curve, $\kappa_{\rm crit}(\lambda)$, 
separating these two phases, as shown in Fig.~\ref{Fig6a}. The critical spin and dimer exponents 
vary continuously on this curve \cite{Lin2012}. 
All our scaling results for the correlators were obtained for parameter values falling on this critical 
curve, and we have improved the estimation of this curve over Ref.~\cite{Lin2012} by going to larger chain lengths. 
Measurements in the valence-bond basis are done by following the mapping to loop estimators developed in \cite{Beach2006}.
To avoid getting locked into winding sectors at large $\lambda$, we have also introduced simultaneous updates for identical 
singlets in the bra and ket confgurations \cite{me} which considerably improves the sampling efficiency.

\begin{figure}[t]
\includegraphics[width=\hsize]{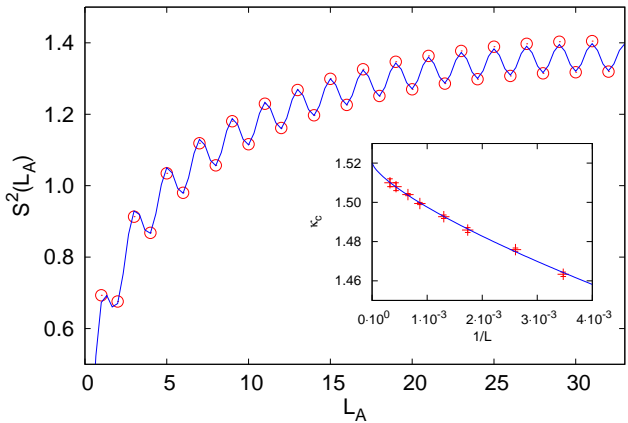}
\caption{EE of the 1D critical APS at $\lambda=1$ for a periodic chain of $L=64$ sites. A fit 
to the form predicted by Eq.~\eqref{ent}, along with finite size corrections \cite{XA} causing the visible
even-odd oscillations, gave $c=1.08(4)$. Error bars are of the order of the symbol size. The inset shows the 
extrapolation of the Binder cumulant crossings points (examples of which are seen in Fig.~\ref{Fig6}), which we
have used to extract the critical $\kappa$ for the infinite system. Here $\kappa_{\rm crit}$ is 1.519(5).}
\label{Fig7}
\end{figure}

We extract crossing points using the Binder cumulant at a particular value of $\lambda$, as shown in Fig.~\ref{Fig6}, and 
estimate the critical dimension of the spin operator by looking at the flow of the exponent with chain length, as described 
in Sec.~\ref{sec3a} in the case of the classical Ising model. We also see that the Binder cumulant becomes negative for 
$\kappa \agt \kappa_{\rm crit}$ and large system sizes (it approaches $0$ for larger $\kappa$). A negative Binder
cumulant is often taken as an indicator of a first-order transition, but for that purpose one also has to check that
the negative peak value grows as the system volume. Rigorous examples exist of more slowly divergent negative
peaks at continuous transitions Ref.~\cite{JinSen}, and, as indicated by our results, the APS may provide yet another 
example where this may be the case.

Next we consider the EE of the APS, using the second Renyi variant $S^2(l_A)$, which is accessible in simulations
using the swap operation and the ratio trick \cite{HM}. We find that the EE follows a profile identical to what we would 
expect from a CFT \cite{Cardy} on a circle (shown for half a periodic chain in Fig.~\ref{Fig7}), i.e., the EE depends on the 
size of the entangled region ($l_A$) as follows:
\begin{equation}\label{ent}
S^2(l_A)=\frac{c}{4}\log\left[\frac{L}{\pi a}\sin \left (\frac{\pi l_A}{L}\right)\right]+d,
\end{equation}
where $L$ is the total system size, and $a$ is the nearest-neighbour distance (set to 1 in our case). 
This form can be derived from the general form for the $n$th Renyi entropy \cite{Cardy}. The constant $d$ is different 
for odd and even sizes and the constant $c$ corresponds to the central charge of the CFT.

\begin{figure}[t]
\includegraphics[width=\hsize]{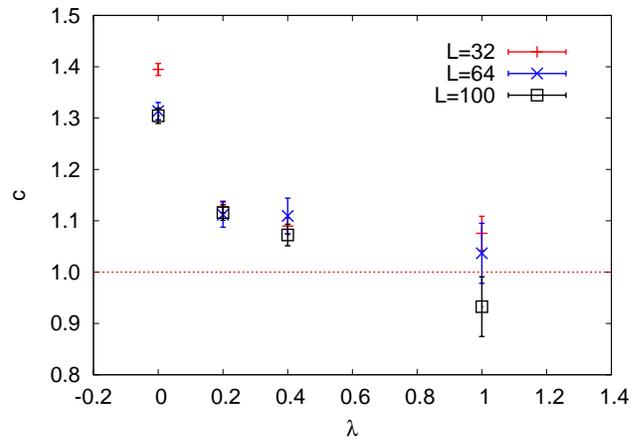}
\caption{EE for the 1D critical APS versus the parameter $\lambda$, extracted as illustrated in
Fig.~\ref{Fig7}.}
\label{Fig8}
\end{figure}

Using this formulation, we can extract the effective central charge of the apparent CFT. We have done this for a set 
of different $\lambda$ (shown in Fig.~\ref{Fig8}) for different sizes and find the effective central charge to be close 
to unity for $\lambda\neq0$. At $\lambda=0$, the VBS phase shifts from being composed of bonds of unit length to one made
out of bonds of length 3 (as unit-length bonds have null weight), and we find the extracted $c$ to be closer to $3/2$ 
in this case. The values $1$ and $3/2$ are both possible CFT central charges in the case of SU($2$) symmetry, and
$c=1$ is associated with continuously varying exponents. Here we have not studied finite-size effects, due to difficulties 
in accurately computing the EE for larger system sizes, and it is not clear whether $c$ will flow with increasing size
to $1$ (in general) and $3/2$ (in the special case of $\lambda=0$), or whether the value changes continuously as we vary 
$\lambda$. In any case, as we will see, the system should not have a CFT description and it is interesting that the
``effective $c$'' nevertheless is close to common CFT values.

\begin{figure}[t]
\includegraphics[width=\hsize]{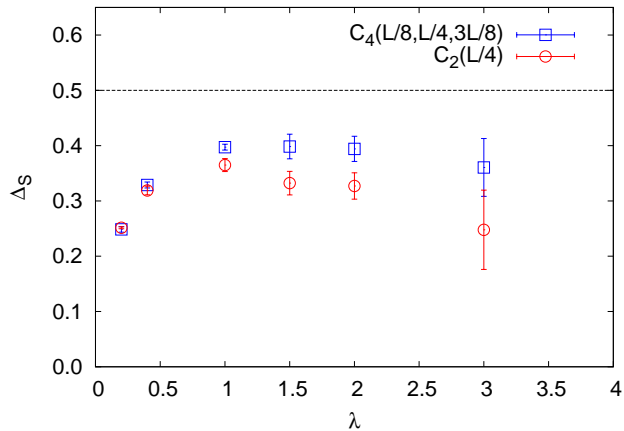}
\caption{Dimension of the spin operator in the 1D critical APS graphed vs the short-bond parameter $\lambda$.
The values were deduced using two- and four-point functions calculated on the critical curve.}
\label{Fig9}
\end{figure}

Next we look at the critical exponent for the spin correlation using the technique of Binder crossings and flowing exponents 
at different $\lambda$. The spin dimension calculated from the two- and four-point functions in Fig.~\ref{Fig9} disagree slightly
when $\lambda \agt 1$, but considering the relatively large error bars we consider the agreement satisfactory nevertheless. The exponent
values differ substantially from the expected CFT value $1/2$ for the entire range of $\lambda$ studied. This rules out a 
$k$=1 WZW description of the critical APS. 

\begin{figure}[t]
\includegraphics[width=\hsize]{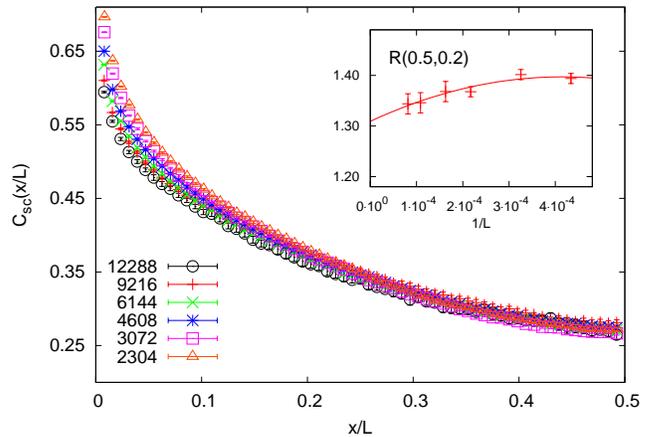}
\caption{Scaled spin correlation functions, defined in Eq.\eqref{sc}, plotted against the separation as a fraction of 
the system size for the critical APS at $\lambda=1$ (where $\kappa_{\rm crit}=1.519$).  The inset shows the ratio $R(0.5,0.2)$ 
Eq.~\eqref{Ratio}, of the correlations at two distances.} 
\label{Fig10}
\end{figure}

We can also study the scaled $r$-dependent correlation function as defined in Eq.~\eqref{sc}, and we expect that it would not 
flow to a constant with respect to point separation with increasing size as there does not appear to be a CFT description for 
this  system. In Fig.~\ref{Fig10} we show results for $\lambda=1$ 
using the scaling dimension $\Lambda_{\sigma}=0.38$ 
from Fig.~\ref{Fig9} and tuning slightly 
to $0.40$ to get data collapse for a large 
range of $\frac{x}{L}$, and find that it does not 
follow the CFT functional form even for a large system of $12288$ sites,
whereas the $J$-$Q$ chain showed this (Fig.~\ref{Fig1}) already for a system of $192$ sites. To check the deviations more
explicitly, we define the ratio of the scaled correlations at two different distances,
\begin{equation}\label{Ratio}
R(r_1,r_2)=\frac{C_{sc}(r_1)}{C_{sc}(r_2)},
\end{equation}
which should approach unity for all $r_1,r_2$ when $L\to \infty$ if the CFT description is valid. In the inset of Fig.~\ref{Fig10} we plot 
$R(0.2,0.5)$ as a function of $1/L$ at criticality and extrapolate it to infinite size using a second-order polynomial. 
We see that the result deviates substantially from unity, suggesting that this indeed is a good CFT test (here with a
negative result). We were not able to look at the behavior of dimer correlators in the APS, as we have done with the $J$-$Q$ 
chain, as the dimer order is quite weak at criticality and not substantial enough to perform a careful finite-size scaling 
analysis with large system sizes.

\section{Conclusion}
\label{sec:concl}.

We have found that the EE of the APS falsely points towards the existence of
a conformal description at criticality. We have determined the inapplicability of a CFT
description by extracting the dimension of the spin operator and by investigating the behavior 
of the two-point spin correlator. As the APS is capable of harbouring a long-range ordered N\'eel 
state in 1D, it is apparent that the Hamiltonian must have long-range interactions on this side of
the phase transition, and it is likely that the effects of these interactions carry on to the critical 
point as well, thus perhaps preventing a CFT description (as that requires sufficient locality of the 
interactions). Note that long-range interactions do not automatically preclude a CFT description and there 
are indeed examples of this, e.g., SU(N) spin chains with exchange interactions decaying with distance $r$ as 
$1/r^2$ \cite{Haldane92}. As the rate of decay is decreased, one would at some point expect the CFT
description to break down.

Examples of direct transitions between antiferromagnetic and VBS phases have already been seen in both frustrated and 
\cite{FQSCLRI} and unfrustrated \cite{LRHAC} spin chains with tunable power-law decaying long-range interactions, and 
it was found that the dynamic 
exponent $z<1$. Thus, space and time do not scale in the same manner and we cannot have a $1+1$-dimensional CFT description of 
such a system (though scale invariance still holds). As the Hamiltonian of the APS is unknown, it is not possible to determine 
$z$, but given that
there is a long-range ordered N\'eel state terminating at the critical curve it seems plausible that the Hamiltonian could
have  $z\not=1$. It might indeed be worthwhile to find the parent Hamiltonian and to test this conjecture.

Our numerics have also shown that the critical APS looks like a CFT when examined only through the lenses of entanglement,
following the predicted scaling with the size of the subsystem of the bipartition and even delivering values of $c$ in the
range of common CFT values (the effective $c$ being close to $1$ and $3/2$). Thus, our study shows that care has to be
taken when using the EE to establish the proper field-theory description, and it is important to also tests the constraints
of CFTs in the details of correlation functions.

\section{Acknowledgements}

We would like to thank Ian Affleck, Roger Melko, and Ronny Thomale for useful discussions.
The computational work reported in this paper was performed in part on the Shared Computing Cluster administered by 
Boston University's Research Computing Services. AWS is supported by the NSF under Grant No.~DMR-1410126.

\bibliography{ref}

\end{document}